\documentclass[twocolumn,showpacs,preprintnumbers,amsmath,
amssymb]{revtex4}
\usepackage{graphicx}
\usepackage{dcolumn}
\usepackage{bm}
\usepackage{tensor}  

\begin{document}
 
\title{Structured light entities, chaos and nonlocal maps.}

\author{A.Yu.Okulov} 
\email{alexey.okulov@gmail.com}
\homepage{https://sites.google.com/site/okulovalexey }
\affiliation{Russian Academy of Sciences, 119991,  
Moscow, Russian Federation.}

\date{\today}

\begin{abstract}
{Spatial chaos as a phenomenon of ultimate complexity 
requires the efficient numerical algorithms. For this 
purpose iterative low-dimensional maps have demonstrated 
high efficiency. Natural generalization of 
Feigenbaum and Ikeda maps may include convolution 
integrals with kernel in a form of Green function of a relevant 
linear physical system. It is shown that such iterative 
$nonlocal$ $nonlinear$ $maps$ are equivalent to ubiquitous 
class of nonlinear partial differential equations 
of Ginzburg-Landau type. With a Green functions relevant 
to generic optical resonators these $nonlocal$ 
$maps$ emulate the basic spatiotemporal phenomena as spatial 
solitons, vortex eigenmodes breathing via relaxation 
oscillations mediated by noise, vortex-vortex and 
vortex-antivortex lattices with periodic 
location of vortex cores. The smooth multimode noise addition 
facilitates the selection of stable entities and elimination 
of numerical artifacts.}
\end{abstract}


\maketitle


\section {Introduction}

Complete spatial synchronization in nonlinear systems 
is replaced 
by turbulent states with a fast decay of correlations. 
The propagation dynamics ranges from switching waves 
to long-living localized excitations and spatiotemporal 
solitons \cite{Cross:1993}. Numerical modeling of these 
complex nonlinear
 distributed systems is based upon finite-difference schemes 
which emulate the basic solutions of partial
 differential equations such as localised solitonic entities 
\cite{Malomed:2006},
propagation fronts, stable phase-locked periodic
 configurations alike lattices and self-organized 
vortex clusters. The alternative numerical approach is outlined. 
It is shown that discrete-time and continuous space dynamical
 systems composed of sequence of local nonlinear point-to-point
Feigenbaum and Ikeda-like maps \cite{Kaneko:1992} and nonlocal
 diffusion-dispersion integral transforms 
are equivalent in a quite general set of cases 
 to conventional partial
 differential equations of Ginzburg-Landau type. Such $nonlocal$
 $maps$ are itself the robust 
numerical schemes that exhibit the transitions from purely chaotic
 states to localization in momentum space inherent to corresponding
 coherent entities alike spatial solitons and vortex 
lattices in the presence of spatial noise. In some experimentally 
achievable cases the statistics 
generated by iterative maps is shown to have the quantitative 
similarity with statistics of multimode random fields 
\cite{Okulov:2009}.

This computational approach is relevant to spatially 
distributed nonlinear systems which demonstrate the  
noise-mediated origin of complex patterns. Superconductors 
of type II in external magnetic field demonstrate 
the hexagonal supercurrent lattices with nodal lines 
of penetrating magnetic induction $\vec {\bf B}$
\cite {Abrikosov:1957}. Thin slice 
nonlinear optical materials with feedback mirrors generate 
triangular intensity patterns \cite {Firth:1991}. 
Wide aperture solid state lasers emit coherent 
multivortex phase-locked beams of rectangular symmetry 
\cite{Chen:2001,Staliunas:1995,Okulov:2008}. Delayed feedback 
systems of coupled oscillators perform reservoir computations 
\cite{Gauthier:2015} 
with Boolean ring networks \cite{Gauthier:2017}. 
 
The transitions from regular patterns to  
turbulent states and vice versa are controlled by a limited 
number of control parameters namely ambient 
temperature ${\bf T}$, magnetic induction, density of carriers 
$n$, gain $G$ and loss $\gamma$, quadratic-qubic-quintic 
nonlinearities $\chi_2,\chi_{3}, \chi_{5}$, carrier frequency 
$\omega=c k = 2 \pi /\lambda$, 
wavelength $\lambda$, dielectric permittivity $\epsilon$ 
, magnetic 
permeability $\mu$ and their spatial distributions.  
Geometrical control parameters are system 
dimensions at different scales, aspect ratios 
(degree of system asymmetry), say ratio of transverse 
dimensions $d,D$ to longitudinal extension $L$, focal 
lengths of optical components $F$,  
frame of reference parameters namely vectors of 
displacement velocity $\vec {\bf V}$ 
and reference frame rotation angular 
velocity $\vec {\bf \Omega_{\oplus}}$ \cite{Okulov:2013}. 
The variations of control parameters force 
system to switch between ordered structures of 
different symmetry or toggle system into spatially 
turbulent states with fast decay of correlations. The 
regions of stability are defined  via calculation of 
instability increments (Lyapunov exponents) \cite{Okulov:2000}.

The above mentioned self-organized structures are 
produced in framework of master equations of Ginzburg-
Landau type. These evolution equations contain 
first order derivative of the order parameter $E$ 
over time $t$ and diffusion terms 
$\Delta_{\bot} = \frac {\partial ^2}{\partial x^2}+
\frac {\partial ^2}{\partial y^2}+
\frac {\partial ^2}{\partial z^2}$ 
with second order derivatives over spatial coordinates, 
so the simplest explicit finite difference 
numerical scheme has the form of iterative mapping, 
which mix the field values in adjacent points of 
numerical mesh $z_{m-1}, z_m ,z_{m+1}$ 
separated by interval $\Delta z$ at 
each time step $\Delta t$ as: 

\begin{eqnarray}
\label {explicit nonlinear diffusion map} 
\frac{E_{n+1,m}-E_{n,m}}{\Delta t} = f( E_{n,m})E_{n,m}+
&& \nonumber \\
\chi \frac{E_{n,m+1}+E_{n,m-1}-2 E_{n,m}}{{\Delta z}^2}.
\end{eqnarray}  

Behavior of such discrete dynamical system (DDS) ranges 
from exactly solvable diffusion for purely real $\chi$  
and diffraction for purely imaginary $\chi$ 
towards spatiotemporal instabilities
\cite{Kaneko:2018}, localized structures\cite{Malomed:2011} 
and turbulence. The role of nonlinear term $f( E_{n,m})$ 
in pattern 
formation is crucial \cite{Cross:1993,Anishchenko:2018}. 
In many cases the numerical 
schemes (\ref {explicit nonlinear diffusion map}) 
are replaced for much more sophisticated 
DDS of implicit type \cite{Malomed:2007} or else convolution - like 
nonlocal integral transforms alike fast Fourier 
transform (FFT) under a proper choice of spatial 
filtering \cite{Siegman:1975,Okulov:1993,Okulov:1994}:
 
\begin{eqnarray}
\label {explicit nonlinear convolution map} 
 E_{n+1}(\vec r)=\int^\infty_{-\infty} K(\vec r - \vec r^{,})
f( E_{n}(\vec r^{,})) d^3{\vec r^{,}} \approx
&& \nonumber \\
\sum_{mx,my,mz} K(\vec r - \vec r_m) f( E_{n}(\vec r_m))
 S({mx,my,mz})    ,
\end{eqnarray} 

where $S({mx,my,mz}) $ is a finite volume element instead of 
infinitesimal one $d^3 \vec r$,
 kernel $K(\vec r -\vec r^{,})$  is a Green 
function of linear version of  
(\ref {explicit nonlinear diffusion map},
\ref {explicit nonlinear convolution map}), 
i.e. response for delta-function $\delta(\vec r -\vec r^{,})$  
in right  part 
of scalar diffusion equation, reads as \\  
$K(\vec r -\vec r^{,})\approx (\chi/\Delta t)^{-1/2}
exp(-\sqrt \chi |\vec r -\vec r^{,}|^2 /\Delta t)$  \\
or for scalar parabolic diffraction equation this reads as\\
$K(\vec r -\vec r^{,})\approx (\chi/\Delta t)^{-1/2}
exp(-i \sqrt \chi |\vec r -\vec r^{,}|^2 /\Delta t)$ 
\cite{Okulov:1988}.\\  
Such discrete numerical approach proved to be extremely 
effective for modeling of nonlinear dynamics 
inherent to evolution partial differential equations 
of so-called parabolic type \cite{Konotop:1994}. Apart from 
nonlocal dispersive laplacian there exists a nonlinear 
term $f(E_n)$ which depends on square modulus of field $E$ as 
in Kolmogorov-Petrovskii-Piskounov (KPP) 
equation \cite{Okulov:1988}:

\begin{equation}
\label {KPP nonlinear diffusion} 
\frac{\partial E(z,\vec r,t )}{\partial z} +
\frac {1}{V} \frac{\partial E}{ \partial t} +
\chi \frac{\partial^2 E} { \partial t^2} = f(E) E .
\end{equation}

Here the diffusion term  $\chi=F^2/k^2 D^2$ may be responsible 
for spatial filtering of high transverse harmonics in laser cavity 
by intracavity 
diaphragm and iteratively repeated nonlinear transformation 
of the scalar field $E$  in ring laser with 
intracavity second harmonic generation or Raman 
scattering \cite{Okulov:1986}. 
More realistic models are based upon 
nonlinear Shrodinger equation, known also as 
Ginzburg-Landau equation (NLS - GLE) which captures the 
interplay of phase-amplitude modulation during 
propagation of complex field  which is the source of 
modulational instability \cite{Konotop:1994}, 
solitons \cite{Malomed:2006} and collapse \cite{Cross:1993}. 
For propagation of light pulse in 
Kerr dielectric NLS - GLE reads as:

\begin{equation}
\label {GLE} 
\frac{\partial E(z,\vec r,t )}{\partial z} +
\frac{n}{c} \frac{\partial E}{ \partial t} +
\frac {i}{2k} \Delta_{\bot} E = i k n_2 |E|^2 E .
\end{equation}

The similar dynamics is embedded in Gross-Pitaevskii 
equation for macroscopic wavefunction $\Psi$ of 
Bose-Einstein condensates \cite{Pitaevskii:1999}:

\begin{equation}
\label {GPE} 
i\hbar \frac{\partial \Psi(z,\vec r,t )}{\partial t}=
-\frac{\hbar^2}{2m} \Delta_{\bot} \Psi + U(\vec r,t)\Psi +
\frac{4 \pi \hbar^2 a_S}{m} |\Psi|^2 \Psi .
\end{equation}
 
where $U(\vec r,t)$ is confining potential of arbitrary 
complexity \cite{Lembessis:2018}, 
$a_S$ is scattering length, $m$ is mass of boson. 
In a presence of gain $G$ and losses $\gamma$ the NLS-GLE may 
have  a form of NLS with Frantz-Nodvik resonant gain 
term \cite{Okulov_self:1988}, relevant to 
amplification with stimulated 
emission cross-section $\sigma$ of light pulse 
of duration $T_2<\tau < T_1$ in a rare earth doped dielectric with 
transverse $T_2$ and longitudinal relaxation times $T_1$: 

\begin{eqnarray}
\label {NLS_FN}
\frac{\partial E(z,\vec r,t)}{\partial z} +
\frac{n}{c} \frac{\partial E}{\partial t} +
\frac{i}{2k} \Delta_{\bot} E = i k n_2 |E|^2 E +
&& \nonumber \\
\sigma N_o(z,\vec r) E \exp [-2\sigma \int^t_{-\infty}|E|^2 d t^{'}] 
-\gamma E,
\end{eqnarray}

We will discuss discrete iterative maps 
(\ref {explicit nonlinear convolution map})
for numerical modeling 
of chaotic and regular spatiotemporal 
propagation inherent to equations 
(\ref {KPP nonlinear diffusion},
\ref {GLE},
\ref {GPE},
\ref {NLS_FN}). 
The article is organized as follows: section II is devoted 
to overview of regular and chaotic iterative 
dynamics of real $1\bf D$ \cite {Okulov:1984} 
and complex point  maps \cite{Ikeda:1982}, 
section III outlines the direct link between evolution 
PDE's (3-4) and maps (1-2), section IV contains examples 
of localized solutions obtained with nonlocal 
maps being equivalent to equations (3-4) , 
the spatially periodic and chaotic lattices are obtained 
numerically in 
section V and nonlinear dynamics in the presence of 
multimode fluctuations is presented in section VI 
with conclusive discussion in section VII. 

\section {Iterative maps with universal behavior.}

Feigenbaum demonstrated the universality in iterations 
of real mapping with parabolic maximum of the unit 
interval into itself alike logistic 
map \cite {Feigenbaum:1978}: 

\begin{equation}
\label {logistic map}
E_{n+1}= \lambda_F E_n(1-E_n),
\end{equation}
 	
where the sole control parameter $\lambda_F$ 
completely determines the discrete time evolution 
of this simplest 
dynamical system. In particular he shown that two 
universal irrational numbers, namely $\delta_F=4.6692..$ 
scales the separation of the values of 
$\lambda_F=\lambda_1,\lambda_2,\lambda_3...\lambda_{M-1}.
\lambda_M...$ where period-doubling 
bifurcations occur(fig.1):

\begin{equation}
\label {feigenbaum const}
\delta_F = \lim_{M \rightarrow \infty} 
\frac {{\lambda}_M -{\lambda}_{M-1}}
{{\lambda}_{M-1}-{\lambda}_{M-2}} 
\rightarrow 4.6692,
\end{equation}

and $\alpha_F=2.5...$ scales the 
location of limit cycle points in phase-space.
 
\begin{figure} 
\center
{ \includegraphics[width=8.0 cm]
{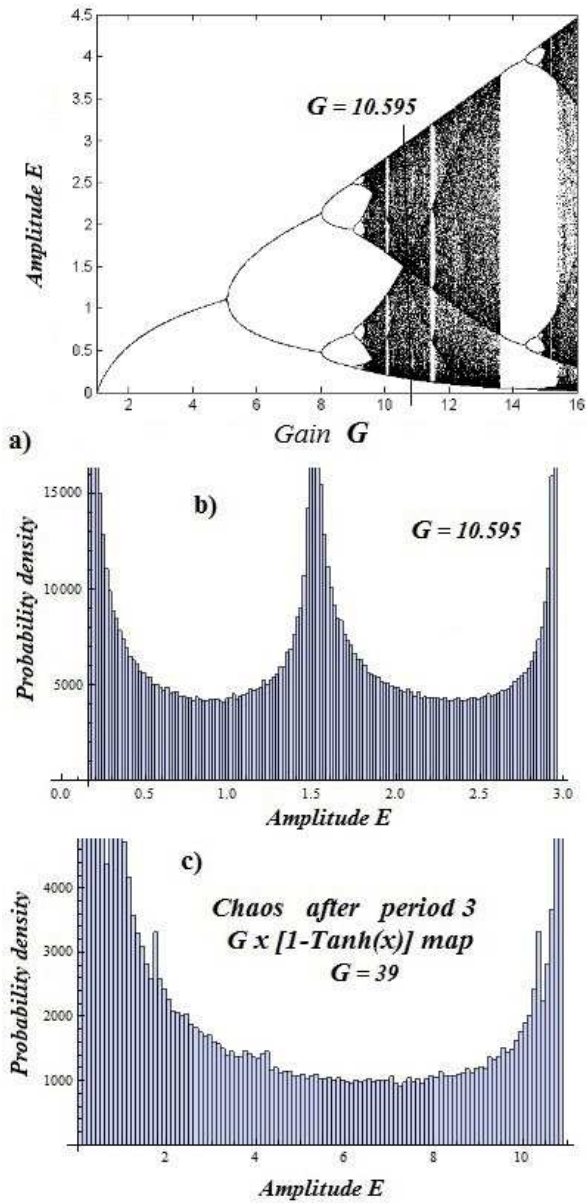}}
\caption{a) Bifurcation diagram of ring laser shows 
distribution of electric field amplitudes $E_n$ at 
gradually increased gain $G=G_{1,2...chaos}$,   
and histograms representing chaotic probability 
densities  $P(E)$ for gain $G=10.595$ (b) and 
$G=39 $ (c) after 5000 iterations.}
\label{fig.1}
\end{figure}

There exists a variety of nonlinear optical systems 
whose dynamics might be approximated by iterates 
of maps with parabolic maxima. The intuitively 
attractive example is a toy model of ring 
laser with intracavity nonlinear losses 
\cite {Okulov:1984} (fig.2a). Radiation 
of electric field amplitude $E_n$ circulates
between confining mirrors along closed trajectory and it passes 
repeatedly through gain element, 
diaphragms  and nonlinear elements.  
The successive passages of field 
through fast amplifying medium with gain 
$E_{n+1}=g(E_n)\rightarrow G E_n {\:}at 
{\:}small{\:} E_n$ 
and nonlinear medium 
with quadratic $\chi_2$ or cubic $\chi_3$ 
susceptibilities are described by following maps:

\begin{equation}
\label {logistic tanh map}
E_{n+1}= g(E_n) \lbrace 1-\tanh [ g(E_n)]\rbrace,
\end{equation}

\begin{equation}
\label {third harmonic map}
E_{n+1}=\frac {g(E_n)}{1+{\chi_3}^2 L^2|g(E_n)|^2},
\end{equation}

\begin{figure} 
\center{ \includegraphics[width=8.0 cm]
{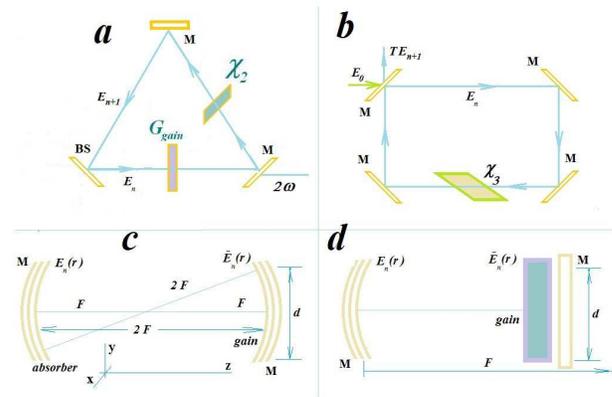}}
\caption{a) Layout of $unidirectional$ single transverse mode 
ring laser with nonlinear losses . 
Envelope of the laser pulse is modulated consecutively from 
one passage to another by hyperbolic tangent 
chaotic map. b) Layout of single transverse mode 
$unidirectional$ ring cavity described by Ikeda map. The phase 
lag between entrance field $E_0$ and intracavity field $E_n$ is 
proportional to light intensity $|E_n|^2$.  c)Layout  
of confocal cavity of length $L_c=2F$ with saturable gain $G(E)$ 
at right mirror and saturable absorber $\alpha(E)$ at the 
opposite one. The fields on opposite 
mirrors $\tilde E_n(\vec r_{bot})$ and $E_n(\vec r_{bot})$ 
are linked via Fourier transform. Spatial soliton 
is formed by transverse modelocking. b) Layout of 
diode-pumped solid-state laser with slightly focusing 
output mirror where vortex-antivortex lattices appear 
due to transverse modelocking 
at high Fresnel numbers $N_f \sim 10^2 - 10^3$.}
\label{fig.2}
\end{figure}
Both maps have 
parabolic maxima and their bifurcation points 
are condensed to 
different values $G=\lambda_{chaos}$ (fig.1) of 
laser system gain 
with the same universal speed $\delta_F=4.6692...$. 
At these bifurcation points 
the deterministic dynamical system generates 
chaotic time series.
Thus in a model of single transversal 
mode solid-state 
laser with nonlinear losses \cite {Okulov:1984} this 
dynamical regime corresponds to generation of 
spatially coherent but temporally chaotic radiation. 
 
Apart from universality the above $1D$ dynamical 
systems might be considered as deterministic source 
of random numbers. In contrast to logistic maps whose 
range of chaotic oscillation amplitudes is 
limited within interval $\lambda \in [0,1)$, 
the time series generated 
by chaotic optical cavities are produced by 
mapping of semi-infinite interval on itself 
\cite{Okulov:1988,Okulov:1986,Okulov:1984}.

Randomization of time series is so strong at 
chaotic accumulation points $\lambda_{chaos}$ that 
probability density functions ($\bf PDF$) for amplitudes $E_n$ 
at a certain values 
of bifurcation parameter $\lambda, G$ are 
very close to 
experimentally obtained histograms for interference 
of statistically independent Stockes pulses 
reflected from independent phase-conjugating 
Brillouin mirrors \cite{Basov_Lett:1980}. 
Noteworthy phases of these 
Stockes pulses are random because stimulated 
Brillouin scattering originates from thermal acoustic 
fluctuations. As a result the recorded 
interference pattern $I \cong 1+\cos(\Delta \phi)/2$ of 
the two beams with phase difference 
distributed uniformly at interval 
$\Delta \phi \in [0,\pi]$ is also random 
though light intensity $I$ has exact 
theoretical probability 
density $P(I)$ \cite{Okulov:1983} (fig.3c): 

\begin{eqnarray}
\label {random intensity Michelson}
P(I) dI \cong \Bigl [1+\cos(\Delta \phi) \Bigr ] 
d (\Delta \phi )  {\:},
&& \nonumber \\
P(I) dI = {\frac {d (\Delta \phi )  } {\pi}}  {\:}, {\:}
P(I) = {\frac{1}{\pi \sqrt{I(1-I)}}} {\:} .
\end{eqnarray}

Both histograms $\bf{(b,c)}$ at fig.3 are perfectly fitted with 
$P(I)=1/[\pi\sqrt{I(1-I)}]$ exact probability densities. 
For this particular case there exists a remarkable coincidence 
of dynamical chaos $\bf PDF$ and interference pattern $\bf PDF$.

\begin{figure} 
\center
{ \includegraphics[width=8.0 cm]
{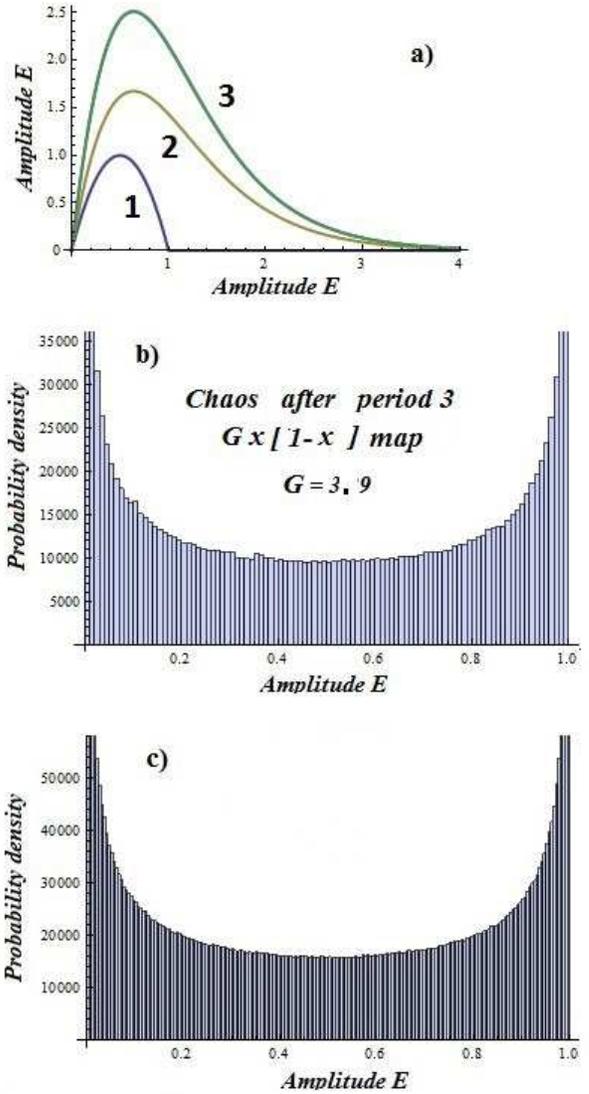}}
\caption{a) Comparison of logistic map ($\bf 1$)
$x \rightarrow  \lambda x(1-x)$ and 
hyperbolic tangent map 
$x \rightarrow  G x[1-tanh(x)]$ ($\bf 2,3$)
for laser with 
nonlinear losses at ($G=6 , 9$), b) 
histogram for full chaos after period 3,
representing chaotic probability 
density  $P(E)$ obtained by 5000 iterations of logistic 
maps with $\lambda=3.9$, c) 
identical probability density 
$P(I)\sim 1/\sqrt{I(1-I)}$
for interference pattern 
for Michelson interferometer 
with independent phase-conjugating mirrors obtained by 
averaging over ensemble of 300 000 counts.}
\label{fig.3}
\end{figure} 
In a more realistic models the nonlinear 
self-phase modulation inherent to the semiconductor 
lasers the Kerr cubic nonlinearity 
[16] may be directly introduced into point 
maps (\ref{logistic tanh map}). 
For this purpose Ikeda constructed a 
map for complex envelopes of electric field 
of ultrashort 
pulses circulating in single transverse mode 
ring cavity \cite{Ikeda:1982} (fig.2b): 
 
\begin{equation}
\label {Ikeda map}
E_{n+1} = R E_{n} \cdot \exp \Bigl [ i k n_o L_c + 
i k n_2 |E_n|^2 L_{nl} \Bigr ]+ T E_0,
\end{equation} 

where $E_0$ is external optical pump, $R$ is reflectivity  
of entrance mirror, $T$ is transmission of entrance mirror, 
$n_0$  is linear index of refraction, $n_2$  is Kerr component of 
nonlinear index of refraction, $L_{nl}$  is width of Kerr slice, 
$L_c$ is length of cavity. 
For more general model of laser cavity with nonstationary 
gain and population inversion lifetime 
$T_1$ the generalized Ikeda map had been introduced in 
\cite{Hollinger:1985} and subsequently generalized for 
wide area laser \cite{Okulov:2008} :
 
\begin{eqnarray}
\label {generalized Ikeda map}
E_{n+1} = T E_0 + R E_{n} \cdot \exp \Bigl 
[i k n_o L_c + \sigma N_n L_{nl} 
i k n_2 |E_n|^2 L_{nl} \Bigr ]  
&& \nonumber \\
\frac {N_{n+1}- N_{n}}{\Delta t}=  +{\frac{N_{0}
-N_{n}}{T_1}}- \sigma N_n |E_n|^2, 
{\:}{\:}{\:} {\:}{\:}{\:} {\:}{\:}{\:} {\:}{\:}{\:} 
\end{eqnarray}
where $N_n$ is population inversion at a 
given bounce of pulse $E_n$  in cavity, $\sigma$ 
is stimulated emission cross-section, $N_0$ 
is the pump rate of amplifying medium, 
$\Delta t= {L_c n }/ c$ is discrete time step of map. 
 
\section {Iterations of discrete maps and parabolic  
partial differential equations.}  

The spatiotemporal evolution of pulse envelops  $E_n$ 
described by equations (9,10,12,13) may be represented by 
sufficiently long operator products as is shown in 
\cite{Okulov:1988}. Indeed each equation (9,10,12,13) in medium 
of length $L$ admits the decomposition for sequence 
of thin slices of linear dispersive elements and 
nonlinear nondispersive elements. 
One may expect that in the limit of infinite 
number  $n \rightarrow \infty$ of an infinitely 
thin slices with  $\Delta L = L /n$ such an 
artificial medium will be equivalent to perfect 
continuous medium. Within 
each slice the propagation of pulse $E_n$ is exactly 
integrable 
so that the following product of maps (2) 
is evident for passage through one slice: 
 
\begin{equation}
\label {Product of maps}
E_{n+1}(\vec r) = \hat D \hat {Fr} f (E_{n}(\vec r)),
\end{equation} 
where 
\begin{equation}
\label {Dispersion operator}
\hat D=1+\frac {\xi \partial^2 }{2 \partial t^2},  {\:}  {\:}  {\:} 
\xi = \Delta z \frac {\partial^2 k}{\partial \omega^2},
\end{equation} 
is dispersion operator, 
\begin{equation}
\label {diffraction operator}
\hat{Fr} = 1+ \frac {i \Delta z} {2k} \Delta_{\bot},
\end{equation} 
is diffraction operator. 
In this picture the 
propagation of pulse  $E_n(\vec r,t)$ 
through $m$ slices is modeled as a product of operators: 
 
\begin{equation}
\label {Long product of maps}
E_{n+m}(\vec r) = \hat D \hat {Fr}  {...}
f \Bigl [  \hat D \hat {Fr} f \Bigl 
[ \hat D \hat {Fr}(f [E_{n}(\vec r))]
\Bigr ] \Bigr ],
\end{equation} 
 for continuous time variable $t=m \Delta t$ this product becomes:
 
\begin{equation}
\label {Continuous time product of maps}
E_{n+m}(\vec r,t) =\lim_{m \rightarrow \infty} 
\hat D \hat {Fr}  {...}
f \Bigl [  \hat D \hat {Fr} 
f \Bigl [ \hat D \hat {Fr}(f [E_{n}(\vec r))]
\Bigr ] \Bigr ],
\end{equation} 

Consider infinitesimal slice $\Delta z = L /n$ and use the 
map (13) for calculation of pulse envelope after passage 
through it: 

\begin{equation}
\label {Product of maps }
E(z+\Delta z,\vec r_{\bot},t) = \hat D \hat {Fr} 
f (E(z,\vec r_{\bot},t)).
\end{equation}

Substitution of operators $\hat D $ 
and $\hat {Fr} $ in this product gives: 
 
\begin{eqnarray}
\label {Product of maps dipersive}
E(z+ \Delta z,\vec r_{\bot},t) = E + 
\Delta z \frac {\partial E}{\partial z}= 
ik {\:}\Delta z{\:} E \Bigl 
[ n_0+n_2 |E|^2 \Bigr ] 
{\:}{\:}
&& \nonumber \\
+E{\:} \Delta z {\:}\frac {\partial^2 k}{\partial \omega^2}
\frac {\partial^2 E}{\partial t^2} +  
\frac {i \Delta z \Delta_{\bot}E }{k}, 
{\:}{\:}{\:}{\:}{\:}{\:}{\:}{\:}{\:}{\:}{\:}{\:}
\end{eqnarray}

where the second identity leads immediately to NLS-GLE equation: 

\begin{equation}
\label {GLE dispersive} 
\frac{\partial E(z,\vec r,t )}{\partial z} +
\frac {\partial^2 k}{\partial \omega^2}
\frac{\partial^2 E}{ \partial t^2} +
\frac {i}{2k} \Delta_{\bot} E + i k \Bigl [
n_0+n_2 |E|^2 \Bigr ] E =0.
\end{equation}

On the other hand the infinite chain of 
operators may be used for construction of exact solution of 
linear Shrodinger equation (NLS with $n_2=0$) 
at finite propagation distance $L=m \Delta z$. 
For this purpose consider pulse propagation in 
free space using the following map:

\begin{equation}
\label {free space diffraction operator}
E(z+\Delta z, \vec r_{\bot}) = \Bigl \lbrace 
1+ \frac {i d z} {2k} \Delta_{\bot} \Bigr 
\rbrace E(z, \vec r_{\bot}),
\end{equation} 

Let us decompose envelope $E$ in Fourier integral: 

\begin{equation}
\label {Fourier initial field}
E(z=0, \vec r_{\bot}) = \frac {1}{ 2\pi} 
\int^{\infty}_{-\infty} 
\exp[i \vec \kappa \cdot \vec r_{\bot}] 
\tilde E (z=0,\vec \kappa) d^2{\vec \kappa}, 
\end{equation} 

The substitution in 
(\ref {free space diffraction operator}) leads to Fourier 
components after first iterate, i.e. after propagation 
distance $\Delta z$:

\begin{equation}
\label {free space diffraction in Fourier space}
E(\Delta z, \vec \kappa) = \Bigl 
\lbrace 1- \frac {i d z \vec \kappa^2} {2k} \Bigr \rbrace 
E(0, \vec \kappa),
\end{equation} 

The next iterates are represented as follows:

\begin{equation}
\label { b free space diffraction in Fourier space}
E(2\Delta z, \vec \kappa) = \Bigl \lbrace 
1- \frac {i d z \vec \kappa^2} {2k} \Bigr \rbrace^2 
E(0, \vec \kappa),
\end{equation} 

\begin{equation}
\label {dz_m free space diffraction in Fourier space}
E(m \Delta z, \vec \kappa) = \Bigl \lbrace 
1- \frac {i d z \vec \kappa^2} {2k} \Bigr \rbrace^m 
E(0, \vec \kappa).
\end{equation} 

In order to obtain solution at finite distance $L$  
let us use limit $m \rightarrow \infty$ under 
apparent constraint $m \Delta z = L$ : 

\begin{equation}
\label {L dist free space diffraction in Fourier space}
E(L, \vec \kappa) = \lim_{m \rightarrow \infty} 
\Bigl \lbrace 
1- \frac {i L \vec \kappa^2} {m 2k}  \Bigr \rbrace^m 
E(0, \vec \kappa).
\end{equation} 
After rearrangement of this formula we have:
 
\begin{equation}
\label {re L dist free space diffraction in Fourier space}
E(L, \vec \kappa) = \lim_{\xi \rightarrow \infty}
\Bigl [ \lbrace 1- \frac {1} {\xi}  \rbrace^{\xi} 
\Bigr]^{\frac {iL \vec \kappa^2}{2k}}
E(0, \vec \kappa), \xi=\frac {2k m}{i L \vec \kappa^2} .
\end{equation}

Next because 
${\lim_{{\xi \rightarrow \infty}}}
( 1- \frac {1} {\xi}  )^{\xi} = e$ 
is known as Euler number, 
we have for Fourier components after propagation at 
finite distance $L$: 

\begin{equation}
\label {Long dist free space diffraction in Fourier space}
E(L, \vec \kappa) = \exp \Bigl [ \frac {iL \vec \kappa^2}{2k} 
\Bigr ] E(0, \vec \kappa), 
\end{equation}

next after return to coordinate space we have: 

\begin{equation}
\label {after L distance Fourier  field}
E(L, \vec r_{\bot}) = \frac {1}{ 2\pi} 
\int^{\infty}_{-\infty} 
\exp \Bigl [ i \vec \kappa \cdot \vec r_{\bot} 
+\frac {i \vec \kappa^2 L}{2k} \Bigr]
\tilde E (0,\vec \kappa)  {\:} d^2{\vec \kappa}, 
\end{equation} 
 
after substitution of initial Fourier spectrum:
\begin{equation}
\label {initial Fourier  field}
E(0, \vec \kappa) = \frac {1}{ 2\pi} 
\int^{\infty}_{-\infty} 
\exp[- i \vec \kappa \cdot \vec r_{\bot} ] {\:}
E (0,\vec r) {\:} d^2{\vec r_{\bot}}, 
\end{equation} 
we obtain exact solution as Fresnel-Kirchoff integral 
\cite {Okulov:1988}: 
 
\begin{eqnarray}
\label {after L Fresnel-Kirchoff integral}
E(L, \vec r_{\bot}) = \frac {ik \exp [ikL]}{ 2\pi L} 
 {\:} \times {\:}  {\:} {\:}  {\:} {\:} {\:} {\:} 
&& \nonumber \\
\int^{\infty}_{-\infty}  {\:} 
\exp \Bigl [\frac {i k |\vec r_{\bot} - 
\vec r_{\bot}^{,}|^2}{2L} \Bigr ] {\:} 
\tilde E (0,\vec r_{\bot}^{,}) {\:} d^2{\vec r_{\bot}^{,}}, 
\end{eqnarray}

\section {Localized wavetrains as fixed points of nonlocal maps}  

Bifurcation diagrams of discrete maps show the location 
of fixed points (fig.1a). One may use constructive 
analogy between fixed points of finite-dimensional maps 
and stationary self-similar solutions of 
evolution PDE's \cite {Moloney:1988}. 
Original idea was to construct 
eigenfunctions of nonlinear resonators with 
Kerr medium from soliton solutions of NLS-GLE 
using boundary conditions \cite {Moloney:1983}. 
The complications of this technique arise from the fact, 
that exact soliton solutions of NLS-GLE are asymptotic 
objects on the whole propagation axis $z$. As a result 
such a generalization of conventional theory of solitons 
to finite space interval bounded by cavity mirrors is not 
trivial. The alternative approach is to 
use Fox-Lee method when diffraction is taken into account 
by calculation of Fresnel-Kirchoff integrals at each 
 round-trip \cite {Okulov:1988}. For the simplest 
Fabry-Perot resonator the mapping of field at 
$n$-th passage into field at  $n+1$-th passage is as 
follows: 
 
\begin{eqnarray}
\label {Nonloc map after L Fresnel-Kirchoff integral}
E_{n+1}(\vec r) = \frac {ik \exp [ikL]}{ 2\pi L} 
\int^{\infty}_{-\infty} 
\exp \Bigl [ \frac {i k |\vec r - \vec r^{,}|^2}{2L} \Bigr]
&& \nonumber \\
f[ E_n (\vec r^{,})] D(\vec r^{,}) d^2{\vec r^{,}}, {\:} {\:} 
E_{n+1} (\vec r)=\hat {Fr}f[ E_n (\vec r)], 
\end{eqnarray}
where $D(\vec r)$  may be a smooth, 
say hypergaussian or even step-like 
Heaviside function $\theta (d - |\vec r|)$ , 
where $d$ is diaphragm width, $L$ is distance between mirrors. 
This product of convolution integral operator $\hat {Fr}$ and
point map $f(E_n (\vec r))$  had been named as 
nonlocal nonlinear map \cite {Okulov:1988}. 

The basic properties of nonlocal maps  are visible 
clearly when different spatial scales are taken 
into account. Here the key parameter is Fresnel 
number $N_f=k D^2/L$ 
of resonator. For the simplest plane-parallel 
Fabry-Perot cavity  $N_f$ is the number of Fresnel 
zones on a given mirror $M_1$ visible from opposite one $M_2$ 
\cite {Born_Wolf:1972}. There are two limits each 
have clear physical meaning. The first one is 
the limit corresponding to geometrical optics i.e. 
$\lambda \rightarrow \infty$ . In this case evolution of 
spatial structure $E_n(\vec r)$ follows to point 
transformations \cite {Okulov:1986}. The other case is 
a single spatial mode limit $N_f \sim 1$ when spatial 
filtering during each passage through the cavity is 
strong enough so $E_n(\vec r)$ has a predefined single 
mode shape, say $\bf {TEM}_{oo}$ whose amplitude 
evolve in time as (8-9) \cite{Okulov:1984} or as 
(11-12) \cite{Okulov:2008} in a presence of 
self-phase modulation.  
The most interesting case is the intermediate one $N_f > 1$
when mode interactions are mediated by 
nonlinearities, diffraction and dispersion. 

Consider first the formation of solitary waves as a 
result of phase-locking of transverse modes. 
Such a localized wavetrain is expected to be the 
eigenfunction of nonlocal map 
$E_{n+1} (\vec r)=\hat {Fr}f[ E_n (\vec r)]= 
\Upsilon E_n (\vec r)$ , where $\Upsilon$ is eigenvalue. 
To get exact solutions for localized transverse 
structures one may consider a specially configured  
nonlinear cavities. In particular the confocal cavities 
(fig.2c) \cite{Okulov:1988,Okulov:2000} which are 
known to have the set of 
degenerate eigenmodes with identical frequencies greatly 
facilitates the phase-locking of eigenmodes 
and formation of stable nonlinear localized 
wavetrains. 
Noteworthy the excitation of threshold solitons due to 
saturable absorption and gain 
\cite{Okulov:2000,Okulov:1999}
had been realized experimentally in confocal 
resonator(fig.2c) \cite {Taranenko:1998}.

The other possibility is the excitation of thresholdless 
spatial solitons due to gain saturation or nonlinear parametric 
processes alike second harmonic generation 
\cite{Okulov:1988}. The cavity is again the confocal 
Fabry-Perot resonator with two spherical mirrors 
$\bf M_1, \bf M_2$ both have the same focal length $F$. 

For detailed numerical modeling it is worthwhile 
consider the saturable gain medium is described by complex 
maps similar to Ikeda equations \cite{Ikeda:1982}: 

\begin{eqnarray}
\label {generalized Ikeda map for resonant gain medium}
E_{n+1} = R E_{n} \cdot \exp \Bigl [i k n_o F + 
i k n_2 |E_n|^2 L_{am} + \sigma_{am} N_n L_{am} \Bigr ]
{\:}{\:} 
&& \nonumber \\
\frac {N_{n+1}- N_{n}}{\Delta t}=  +{\frac {N_{0}
-N_{n}} {{T_1}_{am}}}- 
\sigma_{am} N_n |E_n|^2. {\:}{\:}{\:}{\:}{\:}{\:}{\:}{\:}
\end{eqnarray}

The saturable absorbing medium  described 
by analogous set of Ikeda-like maps: 

\begin{eqnarray}
\label {generalized Ikeda map for resonant absorber medium}
\tilde E_{n+1} = R \tilde E_{n} \cdot \exp \Bigl 
[i k n_o F + 
i k n_2 |\tilde E_n|^2 L_{ab} -
\sigma_{ab} \tilde N_n L_{ab} \Bigr ] , 
&& \nonumber \\
\frac {\tilde N_{n+1}- \tilde N_{n}}{\Delta t}=  
+{\frac {\tilde N_{0}
- \tilde N_{n}} {{T_1}_{ab}}}- 
\sigma_{ab} \tilde N_n |\tilde E_n|^2 
,{\:}{\:}{\:}{\:}{\:}{\:}{\:} 
\end{eqnarray}

where $L_{am}, L_{ab}<<L_r$ are the thicknesses of 
amplifying medium 
and absorber media correspondingly, relevant to 
experimental situation,
$\sigma_{am},\sigma_{ab}$ are the stimulated emission 
cross-sections, ${T_1}_{am},{T_1}_{ab}$ 
are longitudinal 
relaxation times for amplifier and absorber 
correspondingly placed  near opposite confocal mirrors 
or deposited upon their surfaces. The propagation of fields 
$E_{n}$, $\tilde E_{n}$ 
between mirrors is described by already defined 
Fresnel-Kirchoff integral nonlocal maps 
(\ref {Nonloc map after L Fresnel-Kirchoff integral}) with 
parabolic phase-modulation $\exp[-i2k |\vec r|^{2} / F]$ 
induced by mirrors:

\begin{eqnarray}
\label {Nonlocal map for resonant gain-absorber  medium}
E_{n+1}(\vec r) = \frac {ik \exp [ikL]}{ 2\pi L} 
\int^{\infty}_{-\infty} 
\exp \bigl [ \frac {i k |\vec r - \vec r^{,}|^2}{2L}- 
\frac {i 2k |\vec r^{,}|^2 } { F} \bigr ]
&& \nonumber \\
f[ E_n (\vec r^{,})] {\:} {\:}\exp \Bigl 
[-\frac {|\vec r^{,}|^2}{2 d^2} \Bigr ] 
{\:} {\:}d^2{\vec r^{,}},{\:} {\:}{\:} {\:}{\:} {\:}{\:} {\:}
\end{eqnarray}

The standard numerical evaluation with FFT to equilibrium 
stationary  eigenmodes for dicrete time step  
$\delta t=2 L_r n /c$ is achieved for a different levels of 
accuracy via  10 - 150 iterates \cite{Okulov:1994}.
The $2D$ spatial solitons (fig.4) were obtained 
for the cavity (fig.2c). The key assumptions 
for generation of spatial solitons were 
threshold excitation guaranteed via smallness in linear regime 
of gain $\sigma_{amp}  N_n L_{amp}$ compared to absorption 
$\sigma_{ab} \tilde N_n L_{ab}$, faster saturation 
of absorption compared to gain, Kerr self-focusing in gain slice 
and higher saturable self-defocusing in absorber slice. 
Under above restrictions the threshold excitations in 
the form of bright spatial soliton \cite{Malomed:2006} emerge 
after a light $\delta$-like click 
in a given section of computational mesh.

\begin{figure} 
\center{ \includegraphics[width=8.0 cm]
{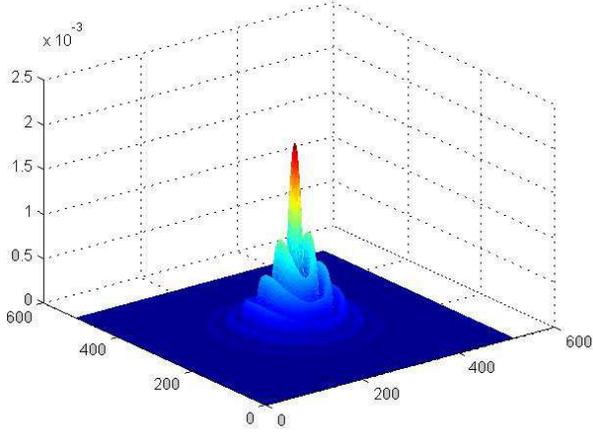}}
\caption{  Spatial soliton  
$E_n(\vec r)$ excited 
in a given area of [512x512] computational mesh 
fitted for confocal cavity fig.2c of length $2F$ with 
saturable gain $G(E)$ at left mirror and 
saturable absorber $\alpha(E)$ at opposite mirror. 
The oscillatory rings around central part of 
soliton subjected to self-phase modulation 
occur due to interference with background.} 
\label{fig.4}
\end{figure}

In order to get analytical solutions the following 
decomposition 
in Taylor series to the third order proved to be useful: 

\begin{eqnarray}
\label {discrete KPP for spatial solitons}
\tilde E_{n+1}(\vec r) = G f [E_{n}(\vec r)]  + 
G  \Bigl ( \frac {4F}{kd} \Bigr )^2 \Delta_{\bot}E_{n}(\vec r),  
G \alpha <1 {\:}{\:}{\:}{\:}{\:}{\:}{\:}{\:}
&& \nonumber \\
f(E_{n})= (\alpha -1)E_{n} \Bigl ( 1-\beta |E_{n}|^2\Bigr )+
E_{n}, {\:}{\:}\beta = \sigma_{ab} {T_1}_{ab}, 
{\:}{\:}{\:}{\:}{\:}{\:}{\:}{\:}{\:}
\end{eqnarray}
where $G=\exp [\sigma_{am} N_0 L_{am}]$.  The exact 
analytical solution $E(x)=E_{n+1}(x) = E_{n}(x)$ in
absorber plane (fig.2c)
in $\bf 1D$ case has the following form: 

\begin{equation}
\label {exact soliton absorber_gain}
 E(x) = A_{ab} 
\sqrt {\frac {2}{\sigma_{ab} {T_1}_{ab} }} sech 
\Bigl [ \frac {x k d}{4 F} \sqrt {\frac {1-\alpha G}{G}} 
\Bigr ], 
\end{equation}
This solution is generalization of the conventional 
Rayleigh formula \cite {Born_Wolf:1972} for focal 
spot size known as $\Delta x \cong \lambda F / d$:
\begin{equation}
\label {rayleigh exact soliton absorber_gain}
 \Delta x \cong \lambda F / d, \Delta x = 
{\frac {\lambda F 4 }{ d 2 \pi}} \sqrt{\frac {G}{1-\alpha G}} . 
\end{equation}
The effective width is not standard Rayleigh one  
$W_{Ral} \cong \lambda F / d $. 
Indeed in our case the dissipative soliton width 
is modified 
in the presence of gain and losses as: 
$W_{sol1}\cong \lambda F \sqrt G/(d \sqrt {1-\alpha G})$. 
As a result the effective width of spatial soliton   
diverges when gain approaches to lasing threshold 
$G \rightarrow \alpha^{-1}$.
The stationary solution $ \tilde E(x)$ in the 
amplifier plane (fig.2c) is Fourier transform of $ E(x)$ : 
\begin{eqnarray}
\label {exact soliton amplifier_plane}
\tilde E(x)= \sqrt {\frac {ik}{2\pi F}}\int^{\infty}_{-\infty} 
E(x^{,}) \exp \Bigl [ \frac {i k x x^{,}}{F} \Bigr ] d x^{,} , 
\tilde E(x) = {\:}{\:}{\:}{\:}{\:}{\:}{\:}{\:}{\:}
&& \nonumber \\
= A_{am} \sqrt {\frac {\pi^2 G}{2 \sigma_{am} {T_1}_{am} 
(1-\alpha G) }} sech 
\Bigl [ \frac {2 \pi x^{,}}{d} \sqrt {\frac {G}{1-\alpha G}} 
\Bigr ], {\:}{\:}{\:}{\:}{\:}{\:}{\:}{\:}{\:}
\end{eqnarray}
Noteworthy both solutions in Fourier conjugated planes 
are hyperbolic 
secants \cite {Okulov:2000} as it follows from exact 
result for the $sech$ spectrum \cite{Ryzhik:2014}. The 
inherent solitonic relation \cite{Moloney:1988} 
between amplitude and 
soliton width known as "area theorem" is embedded here 
in explicit form. 

Linear stability analysis with respect to small 
amplitude spatial harmonics had been realized 
in \cite {Okulov:2000}
following the standard perturbative 
technique \cite {Okulov_self:1988} with linearized 
master equation (\ref {discrete KPP for spatial solitons}): 
\begin{eqnarray}
\label {linearized master equation}
E_{n+1}(x)= E_{n}(x)+i \zeta \psi_{\zeta} , {\:} {\:}{\:} {\:}
\zeta \psi_{\zeta} = -\alpha G \psi_{\zeta} \cdot {\:} {\:}{\:} {\:}{\:} {\:}
&& \nonumber \\
 \Bigl [ 1-3 \sigma_{ab} {T_1}_{ab} 
sech(\nu x)^2  \Bigr ]+ G \Bigl ( \frac {4F}{kd} \Bigr )^2
{\frac {\partial^2 \psi_{\zeta}}{\partial x^2}},
\end{eqnarray}

where $\zeta$ is instability increment, $\psi_{\zeta}$ are 
linear excitation modes in effective potential produced by 
solitons (\ref {exact soliton absorber_gain}),
(\ref {exact soliton amplifier_plane}). 
The spectrum of these infinitesimal excitations 
$\psi_{\zeta}$ consists of the two sets \cite{Okulov:2000}. 
Noteworthy the 
above eigenvalue problem is isomorphic to quantum 
mechanical problem of scattering a particle 
of mass  $m$ on  $sech^2(\nu x)$ potential 
well \cite {Landau:1977}:  
\begin{eqnarray}
\label {linearized master equation2}
\frac {\hbar^2}{2m}
{\frac {\partial^2 \psi_{\zeta}}{\partial x^2}}
- [i \zeta - U_0  {\:} sech^2 (\nu x)]\psi_{\zeta}=0 , 
{\:} {\:}{\:} {\:}
&& \nonumber \\
\nu= \frac {kd}{4F} \sqrt{\frac {1-\alpha G}{G}},
{\:}{\:} {\:}{\:} {\:}{\:} {\:}{\:} {\:}{\:} {\:}{\:} {\:}
\end{eqnarray}
The first set of excitations consists of unbounded 
running plane 
waves $\psi_{\zeta} \cong \exp(\zeta t + ipx)$ 
 with continuum spectrum  $\zeta$. All $\zeta$ in 
this set are negative, thus all unbounded plane wave 
excitations $\psi_{\zeta}$ quench exponentially in time. The 
other set of bounded excitations $\psi_{\zeta}$, 
contains both negative and positive energies $\zeta$
being equal to positive and negative instability 
increments correspondingly: 
\begin{equation} 
\label {lyapunov exponent for spatial soliton}
\zeta = -\alpha G + {\frac {1-\alpha G }{4}}
\Bigl [ {-1-2 n+ \sqrt{ 1+ 
{\frac {24 \alpha G}{1-\alpha G}}}}{\:}{\:}\Bigr ]^2 .
\end{equation}

One may get exact formulas for boundaries of zero $\zeta$
which separate areas of stable increments from 
unstable ones as is shown at fig.5. 
The vital consequence from this linear stability analysis 
is a necessity of specially 
arranged filtering of spatial harmonics $\psi_{\zeta}$, 
belonging to the bounded set of excitations with 
negative spectrum of energies.  Filtering of these positive 
instability increments $\zeta$ will stop the growth 
of excitations and the stability of solitons 
(\ref {exact soliton absorber_gain}),
(\ref {exact soliton amplifier_plane}) is 
guaranteed in this case. 

\begin{figure} 
\center{ \includegraphics[width=8.0 cm]
{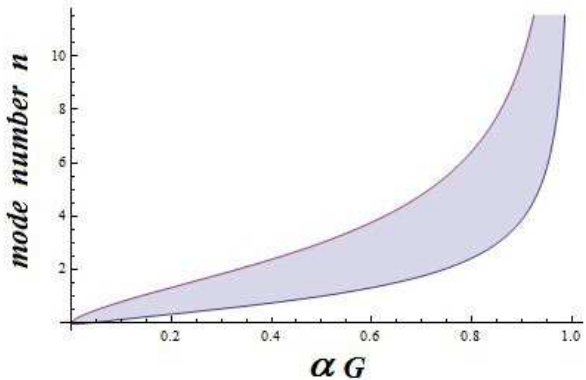}}
\caption{ Location of negative instability increments 
(hatched area) for spatial soliton in 
confocal cavity of length $2F$ with 
saturable gain $G(E)$ at left mirror and 
saturable absorber $\alpha(E)$ at opposite mirror. 
The vertical axis is for number $n$ of spatial harmonic of 
excitation $\psi_{\zeta}$ .} 
\label{fig.5}
\end{figure}

The other confocal cavity configuration is ring 
Sagnac-like cavity with thin-slice nonlinear gain 
medium and spatial filter in Fourier – conjugated planes 
\cite{Okulov:1988} where exact solutions for localized 
solitonic excitations do exist. 
The exact solution for spatial solitons had been obtained by 
searching eigenfunctions of similar nonlocal maps 
which also include Fox-Lee convolution integral:
\begin{equation}
\label {exact soliton Sagnac absorber_gain}
 E(x) = A_{ab} 2 
\sqrt {{\frac {2{G-1}}{\sigma {T_1}_{ab} }}} 
sech \Bigl [{ \frac {x k d \sqrt {G-1}}{4 F}} \Bigr ]. 
\end{equation}
The stability analysis in this case is almost identical 
to the previous one summarized above in eqs. 
(\ref{exact soliton absorber_gain},
\ref{exact soliton amplifier_plane},
\ref{lyapunov exponent for spatial soliton},
\ref{exact soliton Sagnac absorber_gain}) and fig.5. 
In both cavities with gain and losses considered above 
the $1D$ spatial solitons have generic link of width and 
amplitudes in accordance with area theorem. 
The interesting common feature of solitonic widths in both cases:
\begin{eqnarray}
\label {width of solitons}
W_{sol1}= \frac {1}{\nu} = 
{{\frac {4F \sqrt G}{k d \sqrt {1-\alpha G} }}} ; 
W_{sol2}=
{ \frac {4 F} { k d \sqrt {G-1}}} ;
&& \nonumber \\
W_{Ral} \cong 
{1.22 \frac {\lambda F} { d }} ;
{\:}{\:} {\:}{\:} {\:}{\:} {\:}{\:} {\:}{\:} {\:}{\:} {\:}
\end{eqnarray}

is that their widths $W_{sol1}$  and  $W_{sol2}$  are 
basically the generalization of Rayleigh formula for the 
width of the focal spot of a thin parabolic 
lens  $W_{Ral}$ illuminated by a plane monochromatic wave 
of wavelength $\lambda$ and 
aperture $d$ \cite {Born_Wolf:1972}.

\section {Periodic transverse structures in 
laser cavities obtained with nonlocal map 
numerical schemes}
  
Nonlocal maps 
(\ref{Nonloc map after L Fresnel-Kirchoff integral},
\ref{generalized Ikeda map for resonant gain medium},
\ref{generalized Ikeda map for resonant absorber medium})
contain a rich self-organization dynamics. Apart 
from spatial solitons whose shape fit the exact 
solutions (\ref{exact soliton absorber_gain},
\ref{exact soliton amplifier_plane})  
the eigenmodes of wide aperture 
optical resonators with intermediate values of Fresnel 
number $1<N_f<\infty$  demonstrate a various nonlinear 
phase-locking regimes from spatially periodic 
lattices to fully chaotic states which are speckle 
fields characterized by randomly spaced  
optical vortices collocated with zeros of complex field 
amplitudes $E_n(\vec r)$ \cite {Okulov:2009}.  

The simplest case of periodic structure formation 
is possible in a plane-parallel Fabry-Perot cavity 
with a thin slice gain medium having periodic gain 
distribution in transverse directions 
$G(\vec r )=G(\vec r +\vec p)$  , where 
$\vec p=\vec x p_x + \vec y p_y$ , and 
$\vec x, \vec y$,  are unit orts in Cartesian coordinates. 
The self-imaging or Talbot effect is inherent 
to periodic field distributions. 
Because Fox-Lee nonlocal maps (22-24) derived 
above are exact solutions of Maxwell-
Bloch equations in paraxial approximation direct 
substitution of spatially periodic field 
$E_n(\vec r )=E_n(\vec r +\vec p)$  in (22-
24) immediately proves the Talbot identity 
of so-called self-imaging of spatially periodic fields 
at propagation distances $z_T=2 m p^2/ \lambda$, m is 
integer in a set of cases of commensurability of $p_x,p_y$.
The corrections due to finite asymmetric aperture having 
widths $d_x,d_y$  are given by:
\begin{eqnarray}
\label {Talbot self-imaging}
E(x,y,z_T)= \sum_{n_x,n_y} E_{n_x,n_y} 
\exp \Bigl [i 2\pi  \Bigl(\frac {x n_x}{p_x}+
\frac {x n_x}{p_x} \Bigr ) \Bigr]
\times {\:} 
&& \nonumber \\
\exp \Bigl [-\frac {(2 p_x n_x - x)^2}{(1 + iz_T/k d_x^2)2 p_x^2}-
\frac {(2 p_y n_y - y)^2}{(1 + iz_T/k d_y^2)2 p_y^2} \Bigr ]
\times {\:} {\:}
{\:} {\:}
&& \nonumber \\
\Bigl [\Bigl ( 1+\frac {iz_T}{k d_x^2}\Bigr )
\Bigl ( 1+\frac {iz_T}{k d_y^2}\Bigr )\Bigr ]^{-1/2}
\cdot \exp[i k z_T]{\:} .{\:}{\:} {\:}{\:} {\:}
\end{eqnarray}
The generalization of Talbot theorem which is exact
result of conventional diffraction theory to 
nonlinear resonators is as follows \cite {Okulov:1990}. 
It had been shown that for the thin gain slice of arbitrary 
optical nonlinearity the spatially periodic field is 
self-imaged from one mirror upon another 
provided the distance between mirrors is a multiple 
of a Talbot one $L_{c}= m z_T = m 2 p^2/\lambda$ , $m$ is 
positive integer. 
The issue of stability of such periodic configurations 
requires a special attention. 
Nevertheless the numerical investigations with 
Fox-Lee nonlocal maps 
(\ref{Nonloc map after L Fresnel-Kirchoff integral},
\ref{generalized Ikeda map for resonant gain medium},
\ref{generalized Ikeda map for resonant absorber
 medium}) implemented with the aid of standard 
fast Fourier transform (FFT) routines upon 
[128x128] and [1024x1024] computational meshes have shown 
the stable eigenmodes composed of 5x5 and 8x8 
phase-locked periodically spaced filaments 
\cite {Okulov:1993,Okulov:1994} with field 
distributions  $E_{n+1}(\vec r)\cong E_n(\vec r)$ 
close to in-phase wavefronts and 
almost single-lobe far field Fourier spectra with 
suppressed side lobes. 

The self-organized vortex lattices in laser output 
became known since 2001 \cite {Chen:2001} for microchip laser 
oscillators composed from diode-pumped 
thin Nd:YAG gain slice in stable Fabry-Perot 
cavity with long focus $F$ output mirror.
The qualitative agreement with predicted 
periodic vortex lattices \cite{Staliunas:1995} 
had been found. 
The separation of longitudinal modes 
$\Delta \omega_{c}=\pi c/ (n L)$  was large 
enough to facilitate the interaction of a dense set 
of transverse  modes. Fresnel number was in the range  
$10^2< N_f < 10^3$. 
The nonlocal map approach gave straitforward numerical 
technique for 
modeling of the phase-locked vortex lattices (fig.6). 
For accurate numerical simulation the computational 
mesh of a moderate resolution 
[512,512] proved to be sufficient to get the 
stationary vortex-antivortex state.
The detailed pattern of phase dislocations, wavefronts and 
effective fields of velocities of such lattices 
\cite {Okulov:2004} demonstrated the vortex 
pairing (fig.6). 
The effective magnetic fields $\vec  {\tilde B}$ which may 
be realized with such vortex configurations  
\cite {Lembessis:2016,Lembessis:2017}
associated with the each phase-locked vortex 
are counter-directed for 
adjacent vortices. This happens in 
contrast to conventional Abrikosov 
vortex lattices \cite {Abrikosov:1957} where all vortices 
produce co-rotating currents around magnetic field lines.  

\begin{figure} 
\center{ \includegraphics[width=8.0 cm]
{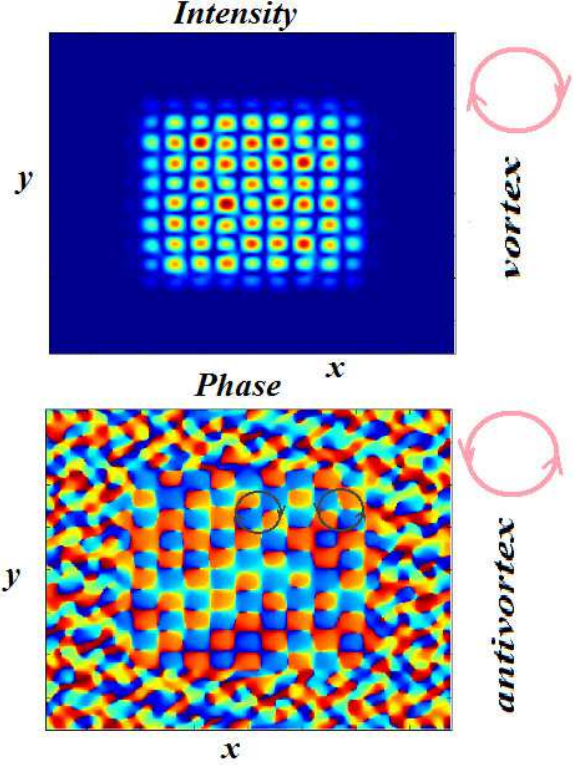}}
\caption{ Vortex lattices with topological 
charge $\ell = \pm 2 $ per each singularity 
in plane-parallel microchip 
laser with high Fresnel number $10^2<N_f<10^3$ obtained 
by virtue of numerical 
modeling with nonlocal maps  
(\ref{Nonloc map after L Fresnel-Kirchoff integral},
\ref{generalized Ikeda map for resonant gain medium},
\ref{generalized Ikeda map for resonant absorber medium}). 
Apart from [7x7] regular lattice around the center of mesh the 
chaotic background of randomly spaced vortex-antivortex pairs 
is seen in bottom "Phase" plot.} 
\label{fig.6}
\end{figure}
Numerical results on melting of this vortex-antivortex 
lattices \cite {Okulov:2003} proved to be in a 
close similarity with 
Berezinskii- Kosterlitz- Thauless scenario 
\cite{V.L.Berezinskii:1971,Kosterlitz:1972}. 
The melting of vortex-antivortex lattices 
ignited by increase of 
optical gain $\sigma_{am} N_0 L_{am}$ was 
envisioned as unbounding the vortices and subsequent 
loss of the long-range order. The observed dynamics 
and proliferation of vortices to random 
locations fits with conventional model 
of speckle fields generated by superposition of plane waves 
with random phases and directions of 
propagation \cite {Okulov:2009}.
 
\section {Modeling of thermal noise mediated patterns}

The inclusion of noise in dynamical equations 
(\ref{Nonloc map after L Fresnel-Kirchoff integral},
\ref{generalized Ikeda map for resonant gain medium},
\ref{generalized Ikeda map for resonant absorber medium})
for multimode laser dynamics is natural, 
because each computational building block of nonlocal 
maps has a clear physical meaning \cite{Anishchenko:2018}. In 
particular the spontaneous emission in cavity 
means the addition to  $E_n(\vec r)$ the multimode speckle field 
composed via superposition of randomly directed plane waves
\cite{Okulov:1991}: 
 
\begin{equation}
\label {multimode random field}
\delta E_n(\vec r)= A_{\delta} \sum_{n_x,n_y}
\exp [ i k_{n_x} x +i k_{n_y} y ], 
\end{equation} 
where $A_{\delta}$ is normalization constant, 
$k_{n_x}^2 +k_{n_y}^2+ k_z^2=k^2 $ . 
The noisy additions to population inversion 
$\delta N_n(\vec r)$   and density of 
carriers $\delta \tilde N_n(\vec r)$   are generated via 
the identical procedure. With these noise sources the 
master equations 
are transformed to somewhat more complicated 
one without decrease of computational 
speed: 

\begin{eqnarray}
\label {noisy_Nonloc map after L Fresnel-Kirchoff integral}
E_{n+1}(\vec r) = \frac {ik_z \exp [ik_z L]}{ 2\pi L} 
\int^{\infty}_{-\infty} 
\exp \Bigl [ \frac {i k_z |\vec r - \vec r^{,}|^2}{2L} \Bigr] 
{\:}{\:}{\:} {\:}{\:}{\:} {\:}{\:}{\:} {\:}
&& \nonumber \\
f[ E_n (\vec r^{,})] D(\vec r^{,}) d^2{\vec r^{,}}+ 
\delta E_n(\vec r), {\:} {\:}  {\:}{\:} {\:} {\:}  {\:}{\:} 
\end{eqnarray}

The equations for thin gain slice are also 
modified straightforwardly as follows: 
 
\begin{eqnarray}
\label {noisy generalized Ikeda map for resonant gain medium}
E_{n+1} = R E_{n} \exp \Bigl 
[i k_z n_o F+ 
i k_z n_2 |E_n|^2 L_{am} + \sigma_{am} N_n L_{am} \Bigr ]
&& \nonumber \\
\frac {N_{n+1}- N_{n}}
{\Delta t}=  +{\frac {N_{0}
-N_{n}+\delta N_n(\vec r)} {{T_1}_{am}}}- 
\sigma_{am} N_n |E_n|^2.{\:} {\:}{\:}{\:} 
{\:}{\:}{\:} {\:}{\:}{\:} {\:}{\:}
\end{eqnarray}

The same holds for the thin saturable absorber slice: 
 
\begin{eqnarray}
\label {noisy generalized Ikeda map for resonant absorber medium}
\tilde E_{n+1} = R \tilde E_{n} \exp \Bigl 
[i k_z n_o F + 
i k_z n_2 |\tilde E_n|^2 L_{ab} + 
\sigma_{ab} \tilde N_n L_{ab} \Bigr ]
&& \nonumber \\
\frac {\tilde N_{n+1}- \tilde N_{n}}
{\Delta t}=  
+{\frac {\tilde N_{0}
- \tilde N_{n}+\delta \tilde N_n(\vec r)} {{T_1}_{ab}}}- 
\sigma_{ab} \tilde N_n |\tilde E_n|^2.{\:} {\:}{\:}{\:} {\:}{\:}
{\:}{\:}{\:}
\end{eqnarray}

\begin{figure} 
\center{ \includegraphics[width=8.0 cm]
{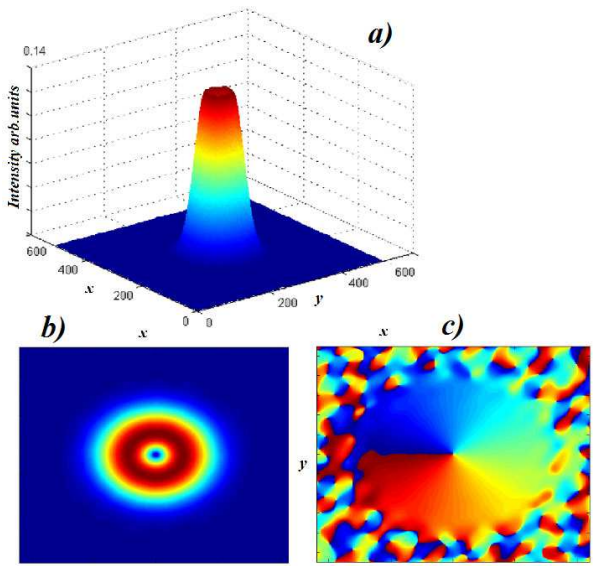}}
\caption{ Vortex with topological charge $\ell=1$ 
in plane-parallel microchip 
laser with low Fresnel number $5<N_f<40$ generated 
via nonlocal Fox-Lee map on [512x512] mesh 
in the presence of 
multimode noise. a),b) are 
intensity $|E_n(\vec r_{\bot})|^2$ plots, c) is phase 
distribution $arg [E_n(\vec r_{\bot})]$ (enlarged).} 
\label{fig.7}
\end{figure} 
 
For low Fresnel number $5 < N_f < 40 $ solid state 
laser cavity the off-axis alignment 
of optical pump may lead to emission of stable 
topologically charged vortices (fig.7) 
\cite {Chen:2018,Chen:2018optexp, Suslov:2013}. Such 
output patterns are highly desirable for 
metrological \cite {Dowling:2012,Dowling:2016} 
and secure free space applications \cite{Zeilinger:2015}. 
The temporal spectrum of such laser oscillator is broadened 
by a set of factors including relaxation 
oscillations \cite {Chen:2001}.  
The fig.8 demonstrates the typical behavior of 
laser output in the presence of noise. 
The stepwise switching on the population inversion 
leads to relaxation oscillation \cite{Letokhov_Suchkov:1967} 
with characteristic 
time scale $\tau_{rel} \cong \sqrt{\tau_{c} T_{1_{am}}}$ 
because of different time scales of $T_{1_{am}}$ 
and photon  lifetime in a cavity $\tau_{c}$(fig.8a). 
As expected in a steady-state regime the power 
spectrum of microchip laser oscillator has a well 
visible relaxation peak due to perpetual disturbances of 
intracavity field $\delta E_n(\vec r)$, population 
inversion $\delta N_n(\vec r)$  and density of carriers 
in absorber $\delta \tilde N_n(\vec r)$ (fig.8b). 
  
\begin{figure} 
\center{ \includegraphics[width=6.0 cm]
{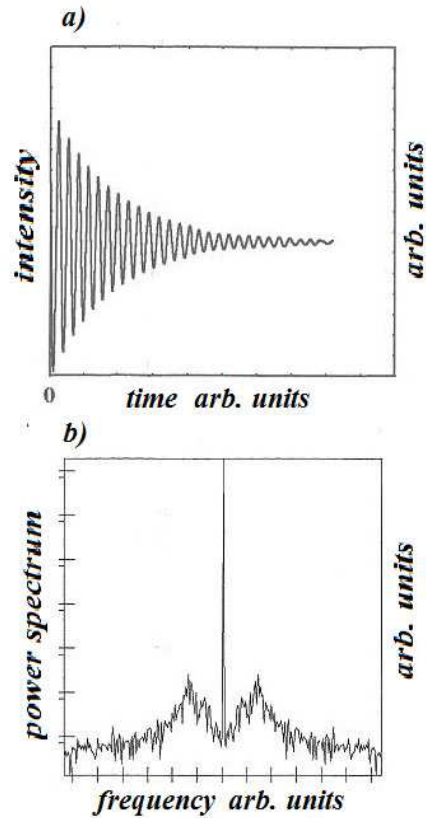}}
\caption{ a)Nonstationary relaxation dynamics of output 
intensity $|E_n|^2$ and  b) power 
spectrum in a model of plane-parallel microchip laser with   
obtained by virtue of numerical modeling with nonlocal 
maps (\ref{noisy_Nonloc map after L Fresnel-Kirchoff integral},
\ref{noisy generalized Ikeda map for resonant gain medium},
\ref{noisy generalized Ikeda map for resonant absorber medium})
shows relaxation oscillation hump.} 
\label{fig.8}
\end{figure}
 
\section {Conclusions}

The toy models of $\bf 1D$ map for unidirectional 
ring laser and $\bf 2D$ Ikeda map for 
standing wave lasers are shown to be easily modified 
into much more realistic models via simple Fresnel-Kirchoff 
convolution integral transform to mediate interaction of 
spatially distributed point maps. The exact and numerical 
solutions for spatial solitons were obtained and their 
stability analysis had been performed. For high Fresnel 
number the iterations of nonlocal maps have shown the 
fast convergence 
rate to stable square vortex lattices known from 
table-top experiments \cite {Chen:2001}. The inclusion of 
noise leads to realistic 
relaxation oscillations power spectra during sufficiently 
short iterations time intervals.

Diversity of dynamical regimes of self-organization in 
spatially distributed nonlinear systems and 
computationally fast generation of stable spatial 
structures via nonlocal maps had been 
demonstrated in this work. The close links with 
conventional evolution equations 
\cite {Malomed:2007,Konotop:1994,Letokhov_Suchkov:1967} 
had been established. The  
stability issues remain still a subject of a very 
complicated analytical studies, as this requires  
the accurate calculation of eigenvalues and 
eigenfunctions of perturbations of solitons, 
vortices and their lattices. The easiness of numerical 
implementation of nonlocal maps especially 
with the aid of FFT routines promise new results in 
this field. In particular the clear physical meaning 
of the Fox-Lee method \cite {Fox_Lee:1966}, 
the natural inclusion 
of boundary conditions in numerical and analytical 
schemes and well elaborated numerical procedures 
for each computational block of nonlocal maps gives  
the firm  guaranties for avoiding numerical 
artifacts, better convergence rates and accurate 
comparison with exact results and alternative numerical 
approaches.  

\bibliographystyle{unsrtnat}

\end{document}